\documentclass[twocolumn,showpacs,preprintnumbers,amsmath,amssymb]{revtex4-1}
\usepackage{graphicx}
\usepackage{textcomp}

\begin{document}
\title{Low-energy doublons in the ac-driven two-species Hubbard model}

  \normalsize
\author{Stefano Longhi and Giuseppe Della Valle}
\address{Dipartimento di Fisica, Politecnico di Milano and Istituto di Fotonica e Nanotecnologie del Consiglio Nazionale delle Ricerche, Piazza L. da Vinci
32, I-20133 Milano, Italy}

%
\bigskip
\begin{abstract}
The hopping dynamics of two fermionic species with different effective masses in the one-dimensional Hubbard model
driven by an external field is theoretically investigated. 
A multiple-time-scale asymptotic analysis of the driven asymmetric Hubbard model  
 shows that a high-frequency bichromatic external field can sustain a new 
 kind of low-energy particle bound state (doublon), in which two fermions of different species occupy nearest neighbor sites and co-tunnel along the lattice. The predictions of the asymptotic analysis are confirmed by direct numerical simulations of the two-particle Hubbard Hamiltonian. 
 \noindent
\end{abstract}



\maketitle

\section{Introduction}
The Hubbard model \cite{Hubbard} is a key theoretical tool in condensed matter physics, which provides crucial insights into electronic and magnetic properties of materials.
Ultracold atomic gases trapped in optical lattices have provided in the past decade a route to simulate the physics  of different kinds of Hubbard models 
originally introduced in the context of condensed-matter
physics \cite{mod1}.  In
addition to exploring the rich equilibrium phase diagram, 
atomic systems can be prepared in highly excited states suitable for
exploring the far from equilibrium dynamics of
strongly correlated systems. The most prominent high-energy
excitations are repulsively bound doubly occupied
sites, called doublons \cite{Win}. Such states were recently observed in
experiments with both bosonic \cite{Win} and fermionic \cite{Winferm}
atoms, and their dynamical properties have been considered in several recent  works (see, for instance, \cite{vari1,vari2,vari3} and references therein).
A doublon is a pair of two fermions tightly bound to
each other. The pair is itinerant; it propagates through the
lattice and thereby acquires a certain energy dispersion. The
pair may decay into its constituents. However, for strongly
repulsive interaction, this decay is suppressed owing to energy conservation.\par
In this work we consider a one-dimensional anisotropic (or asymmetric) Hubbard model (AHM) \cite{AHM1,AHM2,AHM3,AHM4} and show that the application of a high-frequency bichromatic field can  lead to the existence of a new kind of doublons, where two fermions (the 'light' fermion and the 'heavy' fermion) occupy nearest neighbor sites (rather than the same site as in ordinary doublons) and co-tunnel along the lattice.  The AHM was introduced more than 40 decades ago as a model system to describe metal-insulator transitions
in rare-earth materials and transition-metal oxides \cite{AHM1}, in which 'light' and 'heavy' particles  are associated to delocalized Bloch band and localized Wannier states, respectively. The AHM exhibits a rather rich phase diagram, which has been studied in Ref.\cite{AHM2}. The AHM model has gained
a renewed interest in recent years as a simple model to describe binary mixtures of fermionic atoms in optical lattices, in which the two
different fermionic species hop with a different probability amplitude \cite{AHM3}.
Two-species models with different hopping coefficients
can also be realized by trapping atomic clouds with two internal
states of different angular momentum, thereby introducing
a spin-dependent optical lattice with tunable parameters \cite{AHM4}. In the strong interaction regime, doublons of different fermionic species, occupying the same site and tunneling together along the lattice, are found like for the standard Hubbard model. Here we consider the ac-driven AHM, in which an ac external field is applied to the particles. By means of a multiple-time-scale asymptotic analysis of the ac-driven AHM, we show that a  high-frequency bichromatic field can induce a new kind of low-energy doublons, where the light and heavy  fermions occupy nearest-neighbor sites (rather than the same site).\par 
The paper is organized as follows. In Sec.II the driven one-dimensional  Hubbard model for two species with different effective masses is briefly reviewed, and the main equations describing the dynamics in Fock space are derived.  In Sec.III a multiple-scale asymptotic analysis of the driven AHM is presented,  highlighting the existence of a new class of low-energy doublons. The predictions of the asymptotic analysis are confirmed in Sec.IV by direct numerical simulations of the two-particle driven AHM. Finally, in Sec.V the main conclusions are outlined.

\section{ac-driven asymmetric Hubbard model: basic model and two-particle dynamics}

The one-dimensional anisotropic Hubbard model with an external driving field is described by the Hamiltonian (see, for instance, \cite{AHM2})
\begin{eqnarray}
\hat{H} & = & -  \sum_{l, \sigma= \uparrow, \downarrow} J_{\sigma} \left( \hat{c}^{\dag}_{l,\sigma} \hat{c}_{l-1,\sigma}+{\rm H.c.}  \right) +U \sum_{l} \hat{n}_{l,\uparrow} \hat{n}_{l,\downarrow} \nonumber \\
&+ & F(t) \sum_l l (\hat{n}_{l,\uparrow}+\hat{n}_{l,\downarrow}).
\end{eqnarray}
where $\hat{c}^{\dag}_{l,\sigma}$ and $\hat{c}_{l,\sigma}$ are the fermionic creation and annihilation operators   of species $\sigma= \uparrow, \downarrow$ at lattice sites $l=0, \pm 1, \pm 2,...$, $J_{\uparrow}$ and $J_{\downarrow}<J_{\uparrow}$ are the hopping rates of the light and heavy  fermionic species,  $U>0$ is the
on-site repulsion energy, $F(t)$ is the external driving force, and $\hat{n}_{l,\sigma}=\hat{c}^{\dag}_{l,\sigma} \hat{c}_{l,\sigma}$ are the particle number operators at lattice site $l$.  The driving ac force $F(t)$ is assumed to be periodic with period $ T=2 \pi / \omega$.\\
 As briefly mentioned in the introduction, the AHM was earlier introduced to describe metal-insulator transition in rare-earth materials and transition-metal oxides \cite{AHM1}. In 
this case $\sigma$ represents two types of spinless fermions: the ÒlightÓ fermions describe electronic delocalized band (Bloch) states, whereas the ÒheavyÓ fermions tend to be localized on
lattice (Wannier) sites. Nowadays,  the AHM can be simulated  by ultracold atoms loaded
in optical lattices \cite{AHM3,AHM4}. Two-species models with different hopping coefficients
can be realized by trapping atomic clouds with two internal
states of different angular momentum, thereby introducing
a spin-dependent optical lattice, which enables to
modify the anisotropy $a=J_{\uparrow} / J_{\downarrow}$
by controlling the depth of the optical
lattice \cite{AHM4}. Another implementation of the AHM is to trap two different
species of fermionic atoms; in this case 
the anisotropy parameter $a$ is given naturally by the ratio of masses. The forcing $F(t)$ can be 
introduced by periodic lattice shaking  (see, for instance, \cite{Ari}). 
\par 
In the absence of the driving force ($F=0$), the AHM exhibits a rich phase diagram, which has been comprehensively investigated in Ref.\cite{AHM2}. 
Here we focus our attention to the dynamics of two fermions of different species, i.e. one light and one heavy fermion, driven by an external ac force $F(t)$, and we wish to highlight the existence of a novel  low-energy particle bound state (doublon) sustained by the external field which does not have any counterpart in the undriven AHM. To this aim, let us indicate by $a_{n,m}(t)$ the amplitude probability to find the fermion of species $\uparrow$ at lattice site $n$ and the fermion of species $\downarrow$ at lattice site $m$, i.e. let us expand the state vector $| \psi(t) \rangle$ of the system in Fock space as 
 \begin{equation}
 | \psi(t) \rangle= \sum_{n,m}a_{n,m}(t) \hat{c}_{n,\uparrow}^{\dag} \hat{c}_{m,\downarrow}^{\dag}|0 \rangle.
 \end{equation}
  The evolution equations for the amplitude probabilities $a_{n,m}$, as obtained from the Schr\"{o}dinger equation $i \partial_t | \psi \rangle = \hat{H} | \psi \rangle$ with $\hbar=1$, read explicitly
\begin{widetext}
 \begin{equation}
 i \frac{da_{n,m}}{dt}  =  - J_{\uparrow}  \left( a_{n+1,m}+a_{n-1,m} \right) -J_{\downarrow} \left( a_{n,m-1}
  + a_{n,m+1} \right)  +\left[ U \delta_{n,m} +F(t) (n+m) \right] a_{n,m}.
   \end{equation}
   \end{widetext}
 In the absence of the external field ($F=0$), the energy spectrum for the two-particle AHM can be determined analytically (see, for instance, \cite{referee1,referee2}).
 The energy spectrum is continuous and comprises two bands. The first band spans the energy interval $(-4J_{m},4J_{m})$, with $J_{m}=(J_{\uparrow}+J_{\downarrow})/2$, and corresponds to scattered states where the two particles are unbounded and delocalized in the lattice.  The other band, which can be partially overlapped with the former one, corresponds to molecular bound states (doublons), where the two particles are bound and undergo correlated tunneling along the lattice.  In the strong interaction regime $U \gg J_{\uparrow, \downarrow}$, the doublon band is a narrow band energetically well separated from the band of unbound particle states. The narrow doublon band describes highly-excited repulsively bound particle states with a heavy mass and energy $\sim U$, hopping on the lattice with an effective hopping rate $J_{eff}=2 J_{\uparrow} J_{\downarrow}/U$.
If two fermions are initially placed at different lattice sites, i.e. for low energy excitations, the narrow band of high-energy doublon states is not excited and the particle dynamics in Fock space can be formally obtained from Eqs.(3) in the limit $U / J_{\uparrow, \downarrow} \rightarrow \infty$ with $a_{n,n}(t)=0$ (hard-core limit).  In this case, in the absence of the driving field bound particle states are not formed, and the two fermions basically undergo uncorrelated tunneling in the lattice with the only constraint imposed by the hard-core limit, i.e. they are not allowed to cross.  As we will show in the next sections, a high-frequency bichromatic driving field can lead to the existence of a new kind of bound particle states, where the two fermions occupy nearest-neighbor sites and co-tunnel along the lattice ({\it field-induced doublons}).  The existence of such field-sustained low-energy doublons can be proven analytically by an asymptotic analysis of the driven AHM in the hard-core and high-frequency limits, and checked by direct numerical simulations of the AHM in the two-particle subspace.
 
\section{Field-sustained low-energy doublons: theoretical anaysis}

\subsection{Multiple-time-scale asymptotic analysis}

In the hard-core limit, the evolution of the two-particle joint probabilities $a_{n,m}$ with $m>n$  and $m<n$, as described by Eqs.(3), are decoupled and the boundary conditions $a_{n,n}(t) \equiv 0$ hold. For the sake of definiteness, let us consider the case where the particle is constrained to hop in the half plane $m>n$, corresponding to the $\sigma= \uparrow$ fermion initially placed on the left side with respect to the $\sigma= \downarrow$ fermion. To capture the effect of a high-frequency driving field $F(t)$, let us introduce the phase transformation $a_{n,m}(t)=b_{n,m}(t) \exp[-i \Phi(t) (n+m)]$, where we have set
\begin{equation}
\Phi(t)=\int_0^t dt' F(t'),
\end{equation}
and let us introduce the scaled time $\tau= \omega t$, where $\omega$ is the frequency of the ac driving field $[ F(t+ 2 \pi / \omega)=F(t)$]. 
The amplitude probabilities $b_{n,m}(\tau)$ then satisfy the following coupled equations
\begin{widetext}
\begin{equation}
i \frac{b_{n,m}}{d \tau}  =  - \epsilon \kappa_1 \left\{ b_{n+1,m} \exp[-i \Phi(\tau)] +   b_{n-1,m} \exp[i \Phi(\tau)] \right\} - \epsilon \kappa_2 \left\{ b_{n,m+1} \exp[-i \Phi(\tau)] +   b_{n,m-1} \exp[i \Phi(\tau)] \right\}  
\end{equation}
\end{widetext}
for $ m \geq n+1$, with the constraint
\begin{equation}
b_{n,n}(\tau) \equiv 0.
\end{equation}
In Eq.(5) we have set
\begin{equation}
\epsilon \kappa_1 \equiv \frac{J_{\uparrow}}{\omega} \; , \;  \epsilon \kappa_2 \equiv \frac{J_{\downarrow}}{\omega}.
\end{equation}
The high-frequency limit $ \omega \gg J_{\uparrow, \downarrow}$ corresponds to the scaling $\kappa_{1,2} \sim 1$ and $ \epsilon \ll 1$.  In this limit, an approximate solution to Eqs.(5) can be obtained as a power series expansion in $\epsilon$ using a multiple-time-scale asymptotic analysis (see, for instance, \cite{LonghiPRB08}). 
Let us look for a solution to Eqs.(5) of the form
\begin{equation}
b_{n,m}(\tau)=b_{n,m}^{(0)}(\tau)+ \epsilon b_{n,m}^{(1)}(\tau)+ \epsilon^2 b_{n,m}^{(2)}(\tau)+...
\end{equation}
and let us introduce the multiple time scales
\begin{equation}
\tau_0=\tau \; , \; \;  \tau_1 = \epsilon \tau \; , \; \; \tau_2= \epsilon^2 \tau \; , ...
\end{equation}
As is well known, the introduction of multiple time scales is needed to remove the appearance of secular growing terms in the asymptotic analysis that would prevent the validity of expansion (8).
Substitution of the Ansatz (8) into Eqs.(5), using the derivative rule $d / d \tau= \partial_{\tau_0}+ \epsilon \partial_{\tau_1}+ \epsilon^2 \partial_{\tau_2}+...$ and after collecting the terms of the same order in $\epsilon$, a hierarchy of equations for successive corrections to $b_{n,m}$ at various orders is obtained. At leading order ($\sim \epsilon^0$) one simply obtains
$\partial_{\tau_0} b_{n,m}^{(0)}=0$, which yields
\begin{equation}
b_{n,m}^{(0)}=B_{n,m} (\tau_1, \tau_2,...)
\end{equation}
where the amplitudes $B_{n,m}$ vary on the slow time scales $\tau_1$, $\tau_2$,... , and $B_{n,n} \equiv 0$.
The equations at higher orders ( $\sim \epsilon ^k$, $ k \geq 1$) have the general form
 \begin{equation}
 i \partial_{\tau_0} b_{n,m}^{(k)}=-i \partial_{\tau_k} B_{n,m}+G_{n,m}^{(k)}\left ( \tau_0; b_{n,m}^{(j <k)} \right) 
 \end{equation} 
 where $G_{n,m}^{(k)}$ depends explicitly on $\tau_0$ and on the solutions $b_{n,m}^{(j)}$ at previous orders $j=0,1,...,k-1$.  In order to avoid the occurrence of secular growing terms in the solution $b_{n,m}^{(k)}$, the following solvability condition must be satisfied
 \begin{equation}
 i \partial_{\tau_k} B_{n,m}= \langle G_{n,m}^{(k)} \rangle
 \end{equation}
 where $\langle ... \rangle$ denote the dc component of the driving term $G_{n,m}^{(k)}$.  Equation (12) determines the evolution of the amplitude $B_{n,m}$ on the slow time scale $\tau_k$; the correction of $b_{n,m}$ at order $k$ can be then calculated as
 \begin{equation}
 b_{n,m}^{(k)} =-i \int_{0}^{\tau_0} d \xi \left( G_{n,m}^{(k)}-\langle G_{n,m}^{(k)} \rangle \right)
 \end{equation}
 In particular, at order $\sim \epsilon$ one has
 \begin{eqnarray}
G_{n,m}^{(1)}  =   - \kappa_1 \left\{ B_{n+1,m} \exp \left[ -i \Phi(\tau_0)  \right]Ê +
B_{n-1,m} \exp \left[ i \Phi(\tau_0) \right]  \nonumber  \right\} \\
 -  \kappa_2 \left\{  B_{n,m+1} \exp \left[ -i \Phi(\tau_0) \right] +
B_{n,m-1} \exp \left[ i \Phi(\tau_0)  \right] \right\}. \;\;\;
\;\;\;\;\;\;
\end{eqnarray}

\begin{figure}
\includegraphics[scale=0.32]{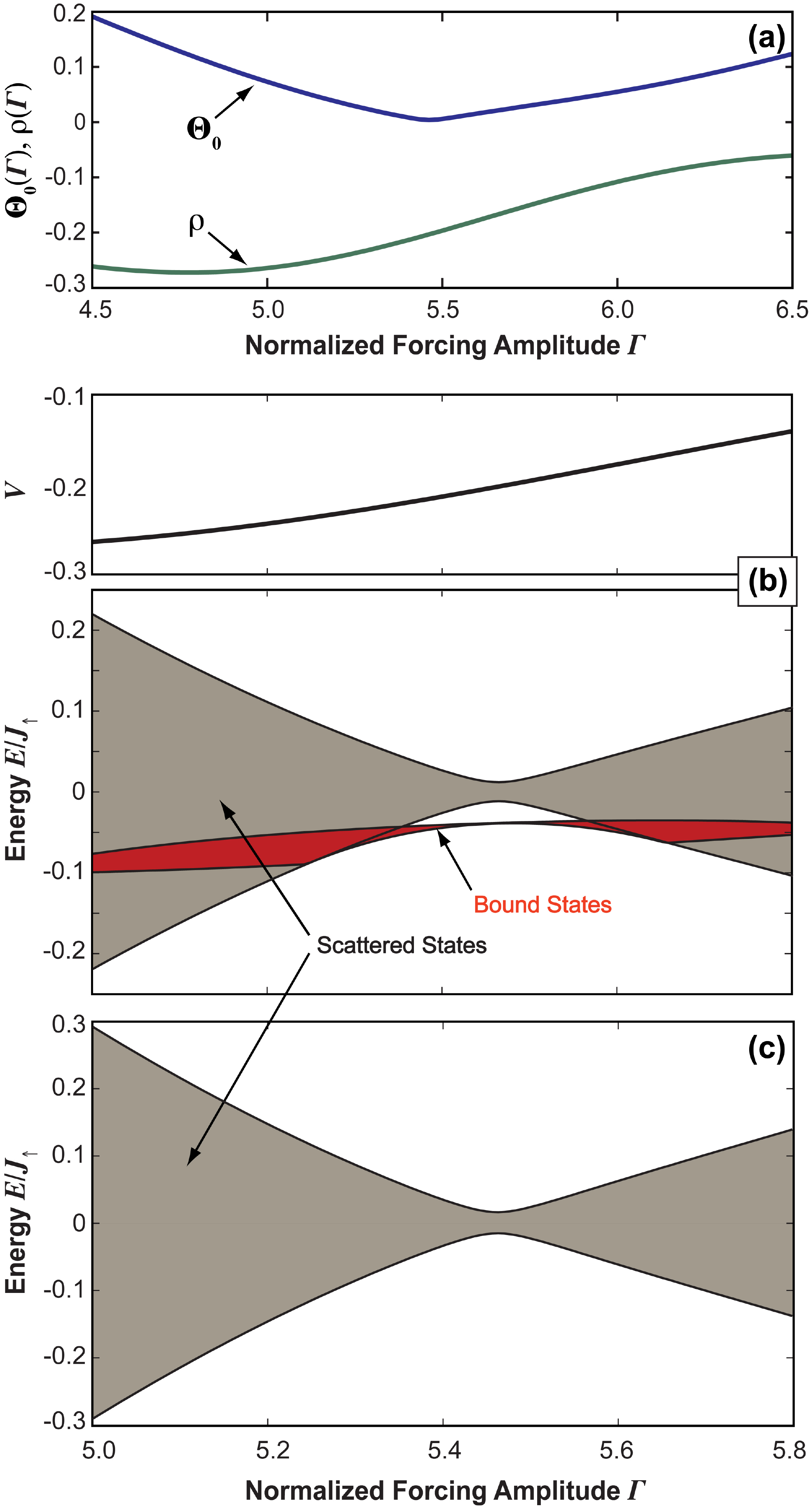}
\caption{(Color online) (a) Behavior of the dimensionless parameters $\Theta_0$ and $\rho$, entering in the asymptotic equations (16) and (17),  versus the normalized amplitude $\Gamma$ of the driving field for $\varphi= \pi/4$. Panel (b) shows the energy spectrum of the asymptotic equations (16) versus $\Gamma$ for $\omega / J_{\uparrow}=4$, $\varphi= \pi/4$, and $a=2$. The behavior of $V$ versus $\Gamma$ is also shown at the top of the figure. The spectrum comprises two distinct bands: the band of unbounded (scattered) particle states, and the band of bound particle states (partially embedded into the former band). (c) Same as (b), but for the isotropic Hubbard model $a=1$. In this case there are not particle bound states.}
\end{figure}
The evolution equation of the amplitude $B_{n,m}$ on the time scale $\tau_1$, as obtained from Eqs.(12) and (14),  corresponds to the well-known application of the rotating-wave approximation to Eqs.(5), where rapidly oscillating terms on the right hand side are neglected. 
To further proceed in the analysis, it is worth introducing the Fourier expansion of the phase term $\exp[i \Phi(\tau_0)]$ by letting
\begin{equation}
\exp[i \Phi(\tau_0)]=\Theta_0+ \sum_{l \neq 0} \Theta_l \exp(i l \tau_0)
\end{equation}
where $\Theta_l$ are the Fourier coefficients.
Note that the condition $\Theta_0=0$ corresponds to the coherent destruction of particle tunneling in the lattice within the rotating-wave approximation \cite{CDT}. In this case, from Eqs.(12), (14) and (15) it follows that the amplitudes $B_{n,m}$ are frozen on the time scale $\tau_1$, i.e. $\partial_{\tau_1} B _{n,m}=0$. However, tunneling is generally allowed at longer time scales, i.e. beyond the crude rotating-wave approximation (see, for instance, \cite{LonghiPRB08,ufff}). Here we assume that $\Theta_0$ is small, of order $\epsilon$, i.e. we assume that the driving force 
parameters are tuned {\it close} to the condition of coherent suppression of tunneling.  The evolution equations of the amplitudes $B_{n,m}$ on the slow time scale $\tau_2$ can be then obtained after some lengthy  but straightforward calculations following the procedure outlined above, once the corrections to $b_{n,m}$ at order $\sim \epsilon$ are calculated using Eqs.(13) and the solvability condition at order $ \sim \epsilon^2$ [Eq.(12)] is explicitly written down.\\ 
If we stop the asymptotic analysis at the order $\epsilon^2$, the temporal evolution of the amplitude probabilities $B_{n,m}(t)$, valid up to the long time scale $\sim 1 / (\omega \epsilon^2)$, is given by $i( d B_{n,m}/dt)= i \omega( \partial_{\tau_0}+ \epsilon \partial_{\tau_1}+ \epsilon^2 \partial_{\tau_2})B_{n,m}=i \omega \epsilon^2 \partial_{\tau_2} B_{n,m}$, which reads explicitly
\begin{eqnarray}
i \frac{dB_{n,m}}{dt} & = & -J_{\uparrow} \left( \Theta_0^* B_{n+1,m} + \Theta_0 B_{n-1,m} \right)  \\
& - & J_{\downarrow} \left( \Theta_0^* B_{n,m+1} + \Theta_0 B_{n,m-1} \right) + V B_{n,m} \delta_{n+1,m}  \nonumber
\end{eqnarray}
for $m \geq n+1$, with $B_{n,n} \equiv 0$. In Eq.(16) we have set
\begin{equation}
V \equiv \frac{\rho}{\omega} (J_{\uparrow}^2-J_{\downarrow}^2)
\end{equation}
where
\begin{equation}
\rho \equiv \sum_{l \neq 0} \frac{| \Theta_l|^2}{l}.
\end{equation}
Equations (16-18) represent the main result of the asymptotic analysis, pushed up to the order $\sim \epsilon^2$, i.e. beyond the most common rotating-wave approximation (the order $\sim \epsilon$).  
\subsection{Two-particle states: field-sustained doublons}
One of the most interesting predictions of the asymptotic equations (16) is that the two-particle energy spectrum comprises, in addition to the Bloch band corresponding to dissociated particles (similar to the case of the undriven AHM in the hard-core limit), an additional  low-energy band corresponding to the two fermions localized in {\it nearest-neighbor} sites that co-tunnell along the lattice.  Such a new kind of doublon states are sustained by the external driving field and, contrary to the doublon states of the undriven (static) HAM \cite{referee1,referee2},  correspond to the two fermions occupying {\it nearest-neighbor} sites.\\
To highlight the physical effects of the external high-frequency field on the dynamics of the two fermions, let us first consider the case of a monochromatic (sinusoidal) driving field, i.e. $F(t)=F_0 \cos (\omega t)$. In this case, $\Theta_l=\mathcal{J}_l(\Gamma)$, where $\Gamma= F_0 / \omega$ and $\mathcal{J}_l$ is the Bessel function of first kind and zero order. Since $| \Theta_{-l}|=| \Theta_l|$, from Eq.(18) it follows that $\rho=0$, and hence $V=0$ in Eqs.(16). In this regime, the field does not sustain doublon states.\\ 
 Let us now consider the case of a driving field with $| \Theta_{-l}| \neq |\Theta_l|$ and $V \neq 0$. Such a condition can be realized, for example, by considering a {\it bichromatic} driving field (see, for instance, \cite{ufff}). In the following analysis, we will specifically consider the following driving force
\begin{equation}
F(t)=F_0 \left[ \cos (\omega t)+ \cos (2 \omega t + \varphi) \right]
\end{equation}
corresponding to a bichromatic field with equal amplitudes $F_0$ for the fundamental and second-harmonic fields and with a phase offset $\varphi$. In this case one has
\begin{equation}
\Theta_l  =  \exp\left(-i \frac{\Gamma}{2} \sin \varphi \right) \sum_n \mathcal{J}_n \left( \frac{\Gamma}{2}\right) \mathcal{J}_{l-2n}(\Gamma) \exp(i n \varphi) \\
\end{equation}
where we have set 
\begin{equation}
\Gamma \equiv \frac{F_0}{\omega}.
\end{equation}
By changing the normalized forcing amplitude $\Gamma$ and the phase offset $\varphi$, the parameters $\Theta_0$ and $V$, entering in Eqs.(16), can be tuned rather arbitrarily. Note that a {\it necessary} condition to have a nonvanishing value of $V$ is that $J_{\uparrow} \neq J_{\downarrow}$, i.e. the original Hubbard model must be {\it anisotropic}. As an example, in Fig.1(a) we show the behavior of $\Theta_0$ and $\rho$ as a function of the normalized forcing amplitude $\Gamma$ for $\varphi=\pi/4$.  Note that at $\Gamma=\Gamma_0 \simeq 5.45$ one has $\Theta_0 \simeq 0$, corresponding to the coherent suppression of particle tunneling, whereas $\rho$ remains finite. Near $\Gamma=\Gamma_0$, the ratio $|V/(J_{\uparrow,\downarrow} \Theta_0)|$ between the energy diagonal defect $V$ and hopping rates $J_{\uparrow,\downarrow} \Theta_0$ can be thus made  large. \\
Following a rather standard procedure (see, for instance, \cite{referee2}), the energy spectrum of Eqs.(16) can be calculated in an exact form; details of the calculations are given in the Appendix. The spectrum is composed by two  bands. The fist one corresponds to unbounded particle states delocalized along the lattice and spans the energy interval $(-4 \kappa, 4 \kappa)$, where $\kappa= |\Theta_0| (J_{\uparrow}+J_{\downarrow})/2$. Interestingly, an additional band appears for a sufficiently large value of $|V|$, namely for $|V|>|\Theta_0| (J_{\uparrow}-J_{\downarrow})$, which corresponds to particle bound states with two fermions occupying nearest neighbor sites and co-hopping along the lattice (field-induced doublons). This band can be partially embedded into the band of unnounded particle states, and its dispersion curve is given in the Appendix.  Figure 1(b) shows, as an example, the energy spectrum of the asymptotic equations (16) versus $\Gamma$ for $a=J_{\uparrow} / J_{\downarrow}=2$,  $\omega / J_{\uparrow}=4$, and $\varphi= \pi/4$, as obtained using Eqs.(A5), (A8) and (A10) given in the Appendix.  The wider band in the figure corresponds to unbounded particle states, whereas the narrower band (partially embedded into the wider one)  corresponds to field-induced doublons.  Note that close to $\Gamma_0$, i.e. close to the coherent destruction of tunneling condition,  the band of doublons is fully outside the band of unpaired states.  Indeed,  in the limit $|V| \gg |J_{\uparrow, \downarrow} \Theta_0|$, the  band of bound particle states turns out to be separated, by $\sim V$, from the band of unbound particle states, and its width is given by $4 |J_{eff}|$, where 
\begin{equation}
J_{eff}=\frac{J_{\uparrow}J_{\downarrow} |\Theta_0|^2}{V}=\frac{J_{\uparrow} J_{\downarrow} |\Theta_0|^2 \omega}{\rho(J_{\uparrow}^2-J_{\downarrow}^2)}
\end{equation}
is the effective hopping rate of the doublon on the lattice. As previously mentioned, the existence of a particle-bound-state band for a driving amplitude $\Gamma$ close to $\Gamma_0$ strictly requires different hopping rates $J_{\uparrow}$ and $J_{\downarrow}$ in the original Hubbard model. As an example, in Fig.1(c) we show the energy spectrum of Eqs.(16) for the same parameter values as in Fig.1(b), except for $a=1$. Note that in this case the energy spectrum comprises a single band, corresponding to unbound particle states.

\begin{figure}
\includegraphics[scale=0.25]{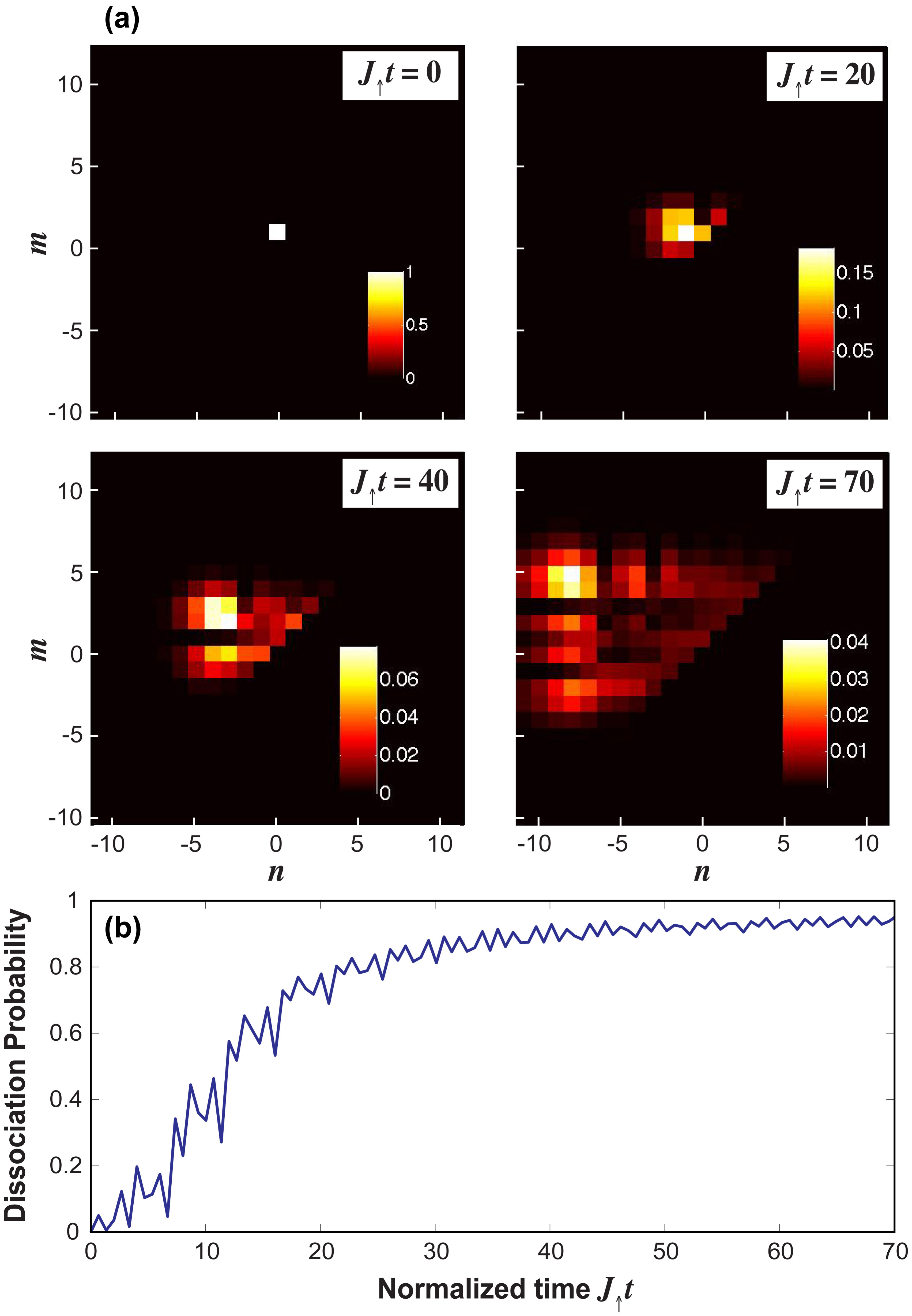}
\caption{(Color online) (a) Evolution of the two-particle joint probability $|a_{n,m}(t)|^2$ at successive times, as obtained by numerical simulations of Eqs.(3), for $a=2$, $ \omega / J_{\uparrow}=4$, $\Gamma=5$, $U/ J_\uparrow=17$, and $\varphi= \pi/4$. The lattice comprises 24 sites. The initial condition is $a_{n,m}(0)=\delta_{n,0} \delta_{m,1}$. In (b) the corresponding evolution of the dissociation probability $P(t)$, defined by Eq.(24), is also depicted.}
\end{figure}
\begin{figure}
\includegraphics[scale=0.25]{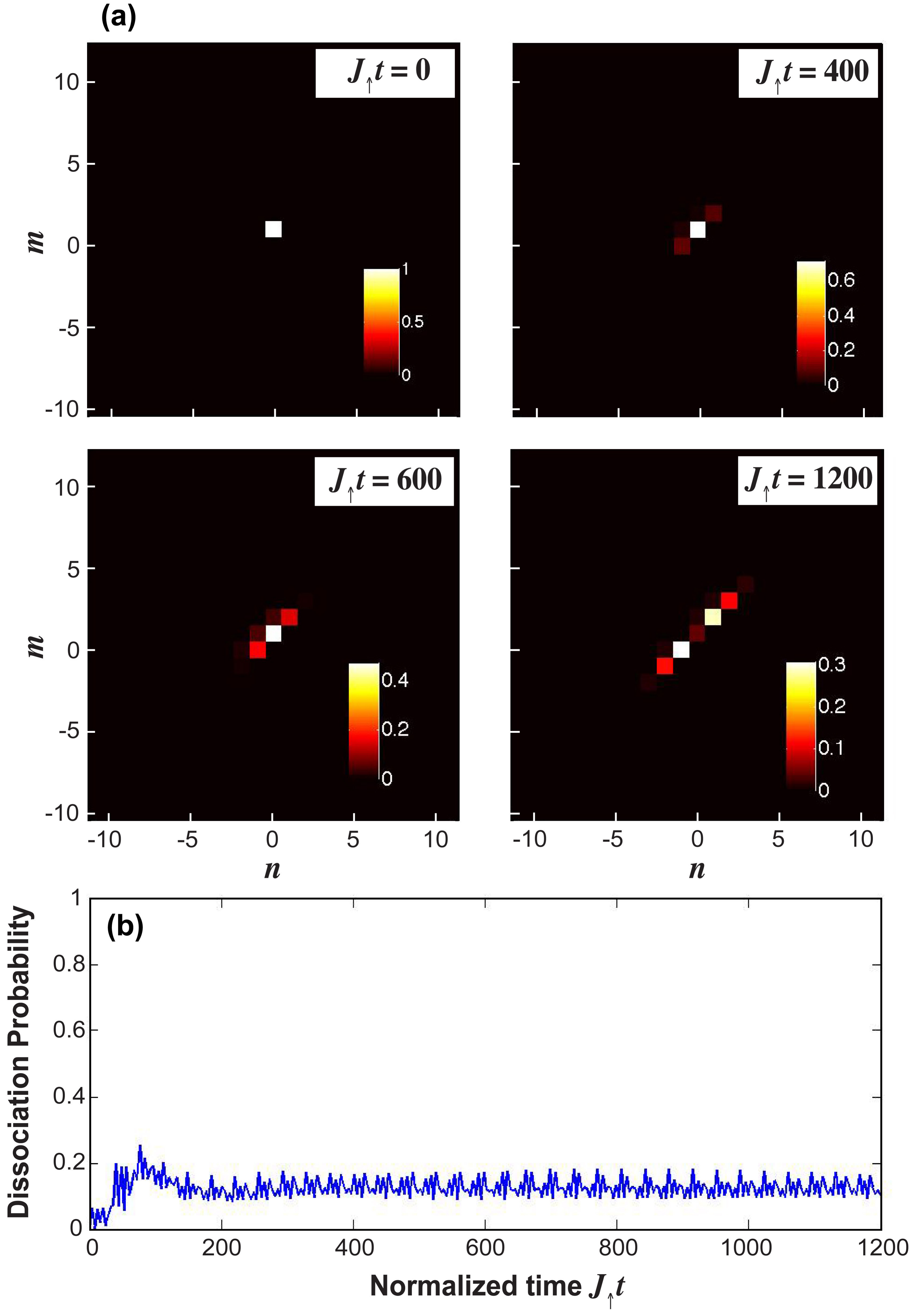}
\caption{(Color online) Same as Fig.2, but for $\Gamma=5.4$.}
\end{figure}
\begin{figure}
\includegraphics[scale=0.25]{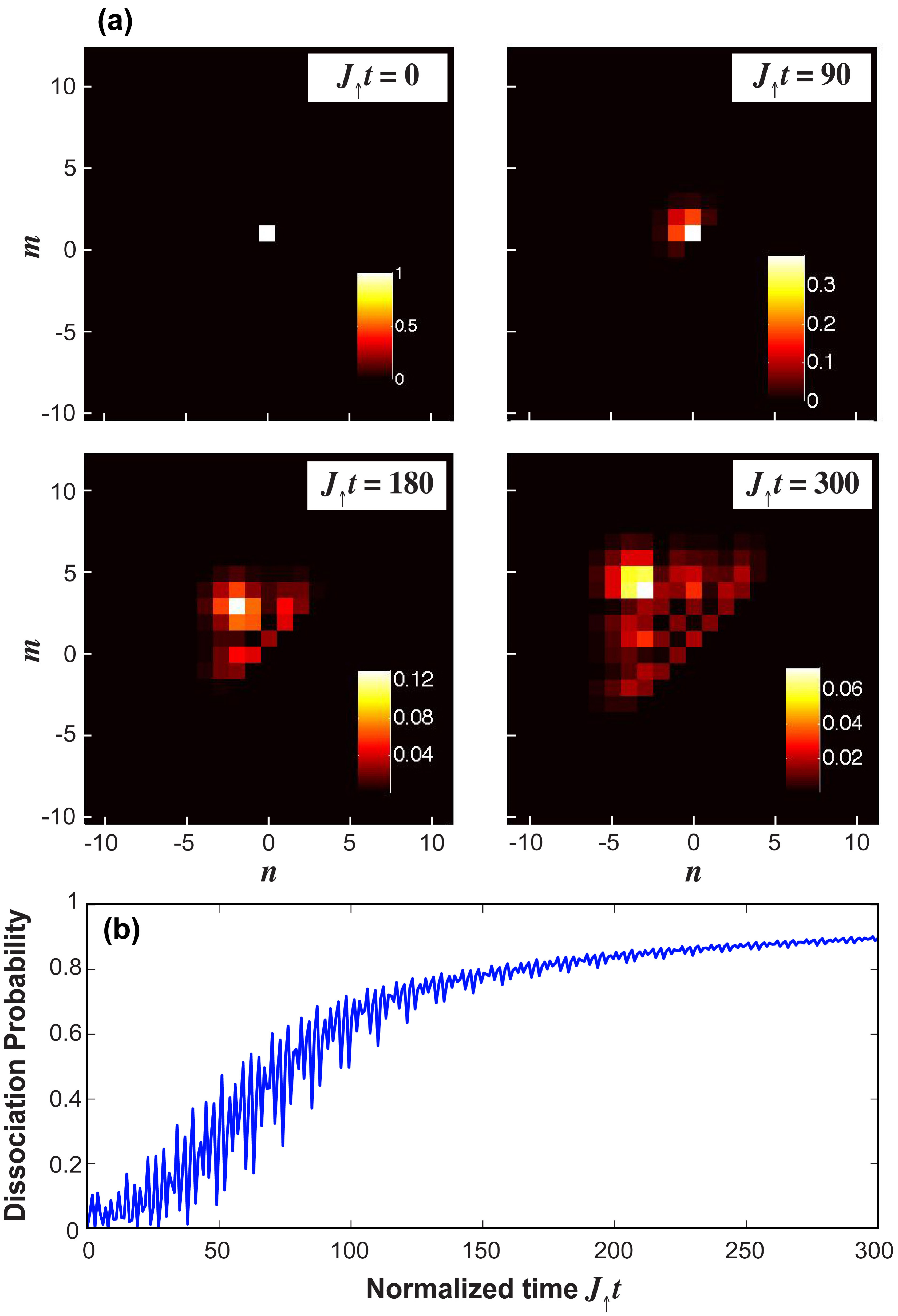}
\caption{(Color online) Same as Fig.3, but for $a=1$.}
\end{figure}

\section{Field-sustained low-energy doublons: numerical results}
 To check the predictions of the asymptotic analysis and the appearance of field-sustained doublon states, we numerically integrated Eqs.(3)  using an accurate fourth-order variable-step Runge-Kutta method in a lattice comprising 24 sites.  Parameter values used in the simulations are $a=J_{\uparrow}/ J_{\downarrow}=2$, $\omega/J_{\uparrow}=4$, and $U/J_{\uparrow}=17$ (hard-core limit). As an initial condition, we assumed that the two fermions  occupy nearest neighbor lattice sites $(0,1)$, namely 
\begin{equation}
a_{n,m}(0)=\delta_{n,0} \delta_{m,1}.
\end{equation}

A bichromatic driving field, defined by Eq.(19), has been assumed with $\varphi= \pi /4$.
Figures 2(a) and  3(a) show the numerically-computed evolution of the joint particle occupation probabilities $|a_{n,m}(t)|^2$ in Fock space at successive times for $\Gamma=5$ and $\Gamma=5.4$, respectively. An inspection of the probability distributions depicted in Figs.2 and 3 clearly show that, as for $\Gamma=5$ the two particles do not bind together and spread along the lattice (Fig.2), for $\Gamma=5.4$  they form a bound state and co-tunnel along the lattice (Fig.3). This is clearly evidenced by the circumstance that in the latter case the two-particle probability distribution remains confined along the diagonal $m=n+1$. The reason thereof is that, according to Fig.1(b), at $\Gamma=5.4$ the band of particle bound states (dublon) detaches from the band of unbounded particle states. Hence, two particle initially placed in nearest-neighbor sites tend to co-tunnel along the lattice remaining in nearest neighbor sites. The probability of the two fermions to co-tunnel along the lattice can be at best captured by plotting, as a function of time, the dissociation probability $P$, defined by
\begin{equation}
P(t)=1-\sum_{n} |a_{n,n+1}(t)|^2.
\end{equation}
The evolution of $P(t)$ is shown in Figs.2(b) and 3(b) for the two normalized driving amplitudes $\Gamma=5$ and $\Gamma=5.4$. Note that, in case of Fig.3(b), the dissociation probability remains small, indicating that the two fermions co-tunnel along the lattice occupying nearest-neighbor sites. It should be noted that field-induced doublon states exist solely for the anisotropic Hubbard model, i.e. for $a \neq 1$, as discussed in the previous section. As an example, in Fig.4 we show the evolution of the joint particle probability distribution $|a_{n,m}(t)|^2$ and of the dissociation probability $P(t)$ for the same parameter values of Fig.3, except for $a=J_{\uparrow} / J_{\downarrow}=1$. Note that, as expected, in this case the two fermions dissociate and spread along the lattice as almost independent particles.

 \section{Conclusions}
 Doublons represent the most prominent high-energy
excitations predicted by the Hubbard model. Such states, which have been recently observed in
experiments with both bosonic \cite{Win} and fermionic \cite{Winferm}
atoms, correspond to repulsive bound particles occupying the same lattice site  that co-tunnel along the lattice. Their dissociation is basically forbidden owing to energy  conservation. In this work we have predicted, both theoretically and numerically, the existence of a new kind of low-energy doublons in the ac-driven anisotropic Hubbard model,
  where a light and a heavy fermion occupy {\it nearest-neighbor} sites and are itinerant in the lattice. Particle binding is here sustained by an external bichromatic driving field, which induces an effective binding energy between  the two fermions.  It is envisaged that our results could stimulate further theoretical and experimental investigations on the physics of ac-driven binary mixtures of fermionic atoms in optical lattices, which provide an experimentally accessible test bed to simulate the anisotropic Hubbard model \cite{AHM3,AHM4}.

\appendix

\section{Energy spectrum of the ac-driven two-particle AHM}
The energy spectrum and corresponding eigenstates of the two-particle states for the ac-driven AHM in the high-frequency and hard-core limits  are obtained from the eigenvalue problem [see Eq.(16) given in the text]
\begin{eqnarray}
E B_{n,m} & = & -J_{\uparrow} \left( \Theta_0^* B_{n+1,m} + \Theta_0 B_{n-1,m} \right)  \\
& - & J_{\downarrow} \left( \Theta_0^* B_{n,m+1} + \Theta_0 B_{n,m-1} \right) + V B_{n,m} \delta_{n+1,m}.  \nonumber
\end{eqnarray}
To determine the energies $E$, we follow a rather standard procedure, in which the
two-body eigenvalue problem (A1) is reduced to a one-body problem (see, for instance, \cite{referee2}). To this aim, let us search for a solution to Eq.(A1) of the form \cite{referee2}
\begin{equation}
B_{n,m}=f(m-n) \exp \left[ -i \beta (m-n) +i (K/2)(m+n) \right]
\end{equation}
where $K$ is the total quasi-momentum of the particles, 
\begin{equation}
{\rm tg} \beta= \frac{J_{\downarrow}-J_{\uparrow}}{J_{\uparrow}+J_{\downarrow}} {\rm tg} ( K/2-\theta)
\end{equation}
and $\theta$ is the phase of $\Theta_0$. Substitution of the Ansatz (A2) into Eq.(A1) yields the following single-particle eigenvalue problem on a semi-infinite one-dimensional lattice for each value of the total quasi-momentum $K$
\begin{equation}
E f(s)= - \sigma(K) [f(s+1)+f(s-1)]+V \delta_{s,1}f(s)
\end{equation}
where $s=m-n \geq 1$ and
\begin{equation}
\sigma(K)=| \Theta_0| \sqrt{J_{\uparrow}^2+J_{\downarrow}^2+2 J_{\uparrow} J_{\downarrow} \cos (K- 2 \theta)}. 
\end{equation}
Equation (A4) is supplemented with the boundary condition $f(0)=0$.  The spectrum and corresponding eigenfunctions of the semi-infinite tight-binding lattice equation (A4) can be readily calculated. \\
I.  {\it Scattered states}. The non-normalizable (scattered) solutions to Eq.(A4) with wave number $q$ are given by
\begin{equation}
f(s)=\exp[i q (s-1)]+r(q) \exp[-iq(s-1)]
\end{equation}
where $r$ is the reflection coefficient, given by
\begin{equation}
r(q)=- \frac{V+ \sigma \exp(-iq)}{V+\sigma \exp(iq)}.
\end{equation}
The corresponding energy $E$ is given by
\begin{equation}
E(K,q)=-2 \sigma (K) \cos q.
\end{equation}
In the original two-particle AHM problem, the scattered states [Eqs. (A2) and (A6)] describe unpaired states, where the two particles are fully delocalized in the lattice. The dispersion curve of the unpaired state band is given by Eq.(A8) and depends on the two quasi-momenta $K$ and $q$ of center of mass and relative motion of the two particles, respectively. Note that the band $E(K,q)$ of unpaired states extends from $-4 \kappa$ to $4 \kappa$, where $\kappa=| \Theta_0| (J_{\uparrow}+J_{\downarrow})/2$. \\
II. {\it Bound states.} For $|V|> \sigma(K)$, Eq.(A4) admits of a bound (normalizable) state, given by
\begin{equation}
f(s)=\exp(- \mu s)
\end{equation}
where $\exp( \mu)=-V/ \sigma(K)$. The corresponding energy is given by
\begin{equation}
E_{doub}(K)=-2 \sigma (K) \cosh \mu = V + \frac{\sigma^2(K)}{V}
\end{equation}
which describes a second band as the total quasi momentum $K$ is varied.
In the original two-particle AHM problem, the eigenstate defined by Eqs.(A2) and (A9) corresponds to a bound particle state (since $|f(s)| \rightarrow 0$ as $s=m-n \rightarrow \infty$), which is delocalized in the lattice (since $B_{n,m}$ is not normalizable).  Equation (A10) thus provides the dispersion relation of the field-induced doublon band. Such band can be partially overlapped with the unpaired band of scattered states, defined by Eq.(A8). In particular, after setting $\sigma_{min}=| \Theta_0| (J_{\uparrow}-J_{\downarrow})$ and $\sigma_{max}=|\Theta_0| (J_{\uparrow}+J_{\downarrow})$, one has:\\
(i) For  $|V|< \sigma_{min}$, there are not bound particle states.\\
(ii) For $\sigma_{min}<|V|< \sigma_{max}$, as $K$ is varied the band of doublon states spans the range $(V+ \sigma_{min}^2/V,2V)$ for $V>0$, or the range $(2V,V+ \sigma_{min}^2/V)$ for $V<0$.\\
(iii) For $|V|> \sigma_{max}$, as $K$ is varied the band of doublon states spans the range $(V+ \sigma_{min}^2/V, V+ \sigma_{max}^2/V)$ for $V>0$, or the range $(V+ \sigma_{max}^2/V, V+ \sigma_{min}^2/V)$ for $V<0$.

\end{document}